\documentclass[showpacs,preprintnumbers,amsmath,amssymb]{revtex4}

\usepackage{rotate}
\usepackage{epsfig}

\def \beq{\begin{equation}}

\def \eeq{\end{equation}}

\def \beqa{\begin{eqnarray}}
\def \eeqa{\end{eqnarray}}
\begin{document}


\begin{titlepage}


\vspace*{1.7truecm}

\begin{center}
{\bf \Large Phenomenological Applications of $k_T$-Factorization \\
\bf --Large Direct CP-Asymmetry in B-meson Decays--}
\vspace*{1.5cm}

{\bf Yong-Yeon Keum}

\vspace*{0.1cm} 

{\it EKEN Lab. Department of Physics \\  
\it Nagoya University, Nagoya 464-8602 JAPAN}

\vspace{0.7truecm}
{\it Email: yykeum@eken.phys.nagoya-u.ac.jp}

\vskip2.0cm
{\bf Abstract}

\vspace*{0.1cm} 

\parbox[t]{\textwidth}{
We discuss applications of the perturbative QCD approach in the exclusive
non-leptonic two body B-meson decays.
We briefly review its ingredients and some important theoretical
issues on the factorization approaches. 
PQCD results are compatible with present experimantal data for the charmless
B-meson decays. We predict the possibility of large direct CP
asymmetry in $B^0 \to \pi^{+}\pi^{-}$ $(23\pm7 \%)$ and
$B^0\to K^{+}\pi^{-}$ $(-17\pm5\%)$. 
We also investigate the Branching ratios, CP asymmetry and isopsin symmetry
breaking in $B \to (K^*/\rho) \gamma$ decays and 
look for the possible new physics
contribution via gluino mediation SUSY  
which can accomodate the large deviation of $S_{\phi K_s}$ from SM.}

\vspace{2.0cm}

\end{center}
{Invited Talk at the 2nd International Conference on Flavour Physics, 
at KIAS, Seoul, Korea  6--11 October 2003}
\end{titlepage}
\thispagestyle{empty}

\setcounter{page}{1}
\normalsize 

\section{Introduction}
Understanding nonleptonic $B$ meson decays is crucial for testing
the standard model(SM), and also for uncovering the trace of new
physics. The simplest case is two-body nonleptonic $B$ meson
decays, for which Bauer, Stech and Wirbel proposed the
factorization assumption (FA) in their pioneering work \cite{BSW}.
Considerable progress, including the generalized FA 
\cite{Cheng94,Cheng96,Soares} and
QCD-improved FA (QCDF) \cite{BBNS}, has been done since this proposal. On
the other hand, technique to analyze hard exclusive hadronic
scattering was developed by Brodsky and Lepage \cite{LB} based on
collinear factorization theorem in perturbative QCD (PQCD). A
modified framework based on $k_T$ factorization theorem has been
given in \cite{BS,LS}, and extended to exclusive $B$ meson decays
in \cite{LY1,CL,YL,CLY}. The infrared finiteness and gauge
invariance of $k_T$ factorization theorem was shown explicitly in
\cite{NL}. Using this so-called PQCD approach, we have
investigated dynamics of nonleptonic $B$ meson decays
\cite{KLS,LUY,KS}. Our observations are summarized as follows:
\begin{enumerate}
\item FA is approximately correct, as our computation shows that
nonfactorizable contributions in charmless $B$ meson decays are
negligible.

\item Penguin amplitudes are enhanced, as the PQCD formalism
inludes dynamics from the region, where the enegy scale $\mu$ runs
to $\sqrt{\bar\Lambda m_b}<m_b/2$, $\bar\Lambda\equiv m_B-m_b$
being the $B$ meson and $b$ quark mass difference.

\item Annihilation diagrams contribute to large short-distance
strong phases through $(S+P)(S-P)$ penguin operators.

\item The sign and magnitude of CP asymmetries in
two-body nonleptonic $B$ meson decays can be calculated, and we
have predicted relatively large CP asymmetries in the $B\to
K^{(*)}\pi$
\cite{KLS,Keum02} and
$\pi\pi$ modes\cite{LUY,KS,Keum01}.
\end{enumerate}

In this talk we summarize shortly ingredient of PQCD method and
important theoretical issues, and show branching ratios of B-meson
decays including $B \to K^{*}\gamma$ decays and possible large
direct CP-violation in $B \to \pi\pi$ and $K\pi$ processes. 
Finally we show a possible solution to explain the large deviation
from SM in the indirect CP asymmetry of $B \to \phi K_s$ mode.
   
\section{Ingredients of PQCD and Theoretical Issues}

{\bf End Point Singularity and Form Factors:} 
If we calculate the $B\to\pi$ form factor $F^{B\pi}$ at large recoil using
the Brodsky-Lepage formalism \cite{Bro,BSH}, a difficulty immediately
occurs. The lowest-order diagram for the hard amplitude is proportional to 
$1/(x_1 x_3^2)$, $x_1$ being the momentum fraction associated with the
spectator quark on the $B$ meson side. If the pion distribution amplitude
vanishes like $x_3$ as $x_3\to 0$ (in the leading-twist, {\it i.e.},
twist-2 case), $F^{B\pi}$ is logarithmically divergent. If the pion
distribution amplitude is a constant as $x_3\to 0$ (in the
next-to-leading-twist, {\it i.e.}, twist-3 case), $F^{B\pi}$ even becomes
linearly divergent. These end-point singularities have also appeared in
the evaluation of the nonfactorizable and annihilation amplitudes in QCDF.

When we include small parton transverse momenta $k_{\perp}$, we have
\begin{equation}
{1 \over x_1\,\, x_3^2 M_B^4} \hspace{10mm} \rightarrow
\hspace{10mm} {1 \over (x_3\, M_B^2 + k_{3\perp}^2) \,\,
[x_1x_3\, M_B^2 + (k_{1\perp} - k_{3\perp})^2]}
\label{eq:4} 
\end{equation}
and the end-point singularity is smeared out 
owing to the Sudakov and threshold resummation effects\cite{KLS}
as shown in figure 1.
\begin{figure}
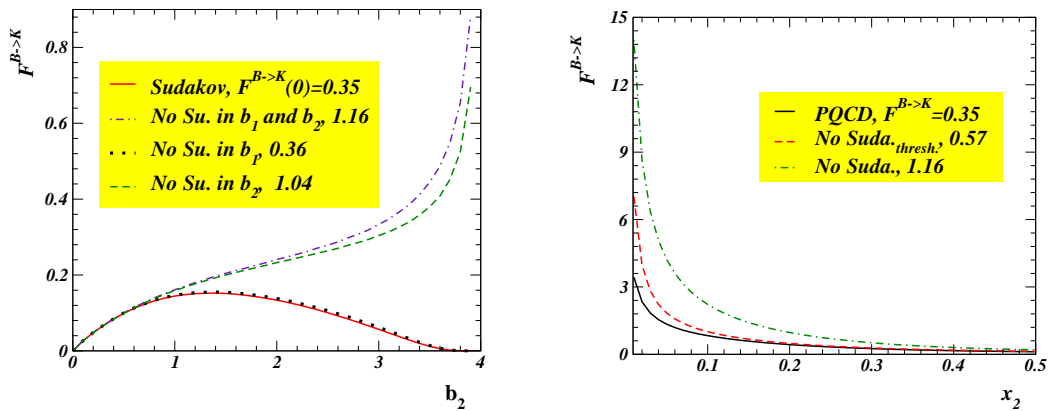

\centerline{\epsfxsize2.5 in \epsffile{fk_b.eps}\hspace{1.0cm}
\epsfxsize2.5 in \epsffile{fk_x.eps} }
\caption{Sudakov suppression and threshold resummation effects in $B \to K$
transition form factor} 
\end{figure}
In PQCD, we can calculate analytically space-like form factors for $B \to P,V$
transition and
also time-like form factors for the annihilation process \cite{CKL:a,Kurimoto}.

\vspace{5mm}
{\bf Strong Phases:} 
While stong phases in FA and QCDF 
come from the Bander-Silverman-Soni (BSS) mechanism\cite{BSS}
and from the final state interaction (FSI), the dominant strong phase in PQCD
come from the factorizable annihilation
diagram\cite{KLS}. In fact,
the two sources of strong phases in the FA
and QCDF approaches are strongly suppressed by the charm mass
threshold and by the end-point behavior of meson wave functions.
So the strong phase in QCDF is almost zero without soft-annihilation
contributions.

\vspace{5mm}
{\bf Dynamical Penguin Enhancement vs Chiral Enhancement:} 
The typical hard scale is about 1.5 GeV as discussed in Ref.\cite{KLS}.
Since the RG evolution of the Wilson coefficients $C_{4,6}(t)$ increase
drastically as $t < M_B/2$, while that of $C_{1,2}(t)$ remain almost
constant, we can get a large enhancement effects from both wilson
coefficents and matrix elements in PQCD. 
 
In general the amplitude can be expressed as
\begin{equation}
Amp \sim [a_{1,2} \,\, \pm \,\, a_4 \,\,
\pm \,\, m_0^{P,V}(\mu) a_6] \,\, \cdot \,\, <K\pi|O|B>
\label{eq:2}
\end{equation}
with the chiral factors $m_0^P(\mu)=m_P^2/[m_1(\mu)+m_2(\mu)]$ for
pseudoscalr meson 
and $m_0^{V}= m_V$ for vector meson.
To accommodate the $B\to K\pi$ data in the factorization and
QCD-factorization approaches, one relies on the chiral enhancement by
increasing the mass $m_0$ to as large values about 3 GeV at $\mu=m_b$ scale.
So two methods accomodate large branching ratios of $B \to K\pi$ and
it is difficult for us to distinguish two different methods in $B \to
PP$ decays. However we can do it in $B \to PV$ because there is no
chiral factor in LCDAs of the vector meson. 

We can test whether dynamical enhancement 
or chiral enhancement is responsible
for the large $B \to K\pi$ branching ratios 
by measuring the $B \to VP,VV$ modes.
In these modes penguin contributions dominate, 
such that their branching ratios are
insensitive to the variation of the unitarity angle $\phi_3$.
Our prediction for various modes are shown at Table 2, in fact,
which is in a good agreement with experimental data.

\vspace{5mm}
{\bf Fat Imaginary Penguin in Annihilation:} 
There is a falklore that annihilation contribution is negligible
compared to W-emission one. In this reason annihilation contribution
was not included in the general factorization approach and the first
paper on QCD-factorization by Beneke et al. \cite{BBNS:99}.
In fact there is a suppression effect for the operators with structure
$(V-A)(V-A)$ because of a mechanism similar to the helicity
suppression for $\pi \to \mu \nu_{\mu}$. However annihilation from 
the operators $O_{5,6,7,8}$ with the structure $(S-P)(S+P)$ via Fiertz
transformation survive under the helicity suppression and can get
large imaginary value. The real part of factorized annihilation contribution
becomes small because there is a cancellation between left-handed
gluon exchanged one and right-handed gluon exchanged one as shown in
Table 1. This mostly pure imaginary value of annihilation is a main
source of large CP asymmetry in $B \to \pi^{+}\pi^{-}$ and $K^{+}\pi^{-}$.
In Table 3 we summarize the CP asymmetry in 
$B \to K(\pi)\pi$ decays.

\section{Numerical Results}
{\bf Branching ratios in Charmless B-decays:} 
The PQCD approach allows us to calculate 
the amplitudes for charmless B-meson decays
in terms of ligh-cone distribution amplitudes upto twist-3. 
We focus on decays
whose branching ratios have already been measured. 
We take allowed ranges of shape parameter for the B-meson wave funtion as 
$\omega_B = 0.36-0.44$ which accomodate to reasonable form factors, 
$F^{B\pi}(0)=0.27-0.33$ and $F^{BK}(0)=0.31-0.40$. 
We use values of chiral factor
with $m_0^{\pi}=1.3 GeV$ and $m_0^{K}=1.7 GeV$.
Finally we obtain branching ratios for $B\to K(\pi)\pi$ 
\cite{KLS,LUY}, 
$K\phi$ \cite{CKL:a,Mishima} $K^{*}\phi$\cite{CKL:b} and 
$K^{*}\pi$\cite{Keum02},
which is well agreed with present experimental data.
\small
\begin{table}[htb]
\caption{Branching ratios of $B \to \pi\pi, K\pi$and $KK$ decays 
with $\phi_3=80^0$, $R_b=\sqrt{\rho^2+\eta^2}=0.38$. 
Here we adopted $m_0^{\pi}=1.3$ GeV,
$m_0^{K}=1.7$ GeV and $0.36<\omega_B<0.44$.
Unit is $10^{-6}$.} 
\label{TABLE1}
\begin{tabular} {|c|ccc|c|c|} \hline 
Modes & CLEO & BELLE & BABAR & ~~~World Av.~~~ & ~~~PQCD~~~  \\
\hline  
$\pi^{+}\pi^{-}$ & $4.5^{+1.4+0.5}_{-1.2-0.4}$ &
 $4.4\pm 0.6 \pm 0.3$ &
 $4.7\pm 0.6 \pm 0.2$ &  
$4.6 \pm 0.4$ &
$5.93-10.99$  \\
$\pi^{+}\pi^{0}$ & $4.6^{+1.8+0.6}_{-1.6-0.7}$ & 
 $5.3 \pm 1.3 \pm0.5$ &
 $5.5^{+1.0}_{-0.9} \pm 0.6$ & $5.3\pm0.8$ & 
$2.72-4.79$    \\ 
$\pi^{0}\pi^{0}$ & $<4.4$ & 
 $<4.4$ &  $<3.6$ & $<3.6$ &
  $0.33-0.65$    \\ 
\hline
$K^{\pm}\pi^{\mp}$ &  
 $18.0^{+2.3+1.2}_{-2.1-0.9}$ &
 $18.5 \pm1.0 \pm0.7$ &  
 $17.9\pm 0.9 \pm 0.7$ &
 $18.2 \pm 0.8$ &  
 $12.67-19.30$    \\ 
$K^{0}\pi^{\mp}$ & 
 $18.8^{+3.7+2.1}_{-3.3-1.8}$ &
 $22.0 \pm1.9 \pm1.1$  &   
 $20.0\pm 1.6 \pm 1.0$ &
 $20.6 \pm 1.4$ &  
 $14.43-26.26$    \\ 
$K^{\pm}\pi^{0}$ &
 $12.9^{+2.4+1.2}_{-2.2-1.1}$ &
 $12.8 \pm1.4^{+1.4}_{-1.0}$ &  
 $12.8^{+1.2}_{-1.0}\pm 1.0$ &
 $12.8 \pm 1.1$ &  
  $7.87-14.21$    \\
$K^{0}\pi^{0}$ &
 $12.8^{+4.0+1.7}_{-3.3-1.4}$ &
 $12.6 \pm2.4 \pm1.4$ &  
 $10.4 \pm 1.5 \pm 1.8$ &
 $11.5 \pm 1.7$ & 
 $7.92-14.27$    \\ 
\hline 
$K^{\pm}K^{\mp}$ &
 $<0.8$ &
 $<0.7$ &  
 $<0.6$ &
 $<0.6$ & 
 $0.06$    \\ 
$K^{\pm}\bar{K}^{0}$ &
 $<3.3$ &
 $<3.4$ &  
 $<2.2$ &
 $<2.2$ & 
 $1.4$    \\ 
$K^{0}\bar{K}^{0}$ &
 $<3.3$ &
 $<3.2$ &  
 $<1.6$ &
 $<1.6$ & 
 $1.4$    \\ 
\hline
\end{tabular}
\end{table} 
\begin{table}[htb]
\caption{Branching ratios of $B \to \phi K^{(*)}$and $K^{*}\pi$ decays 
with $\phi_3=80^0$, $R_b=\sqrt{\rho^2+\eta^2}=0.38$. 
Here we adopted $m_0^{\pi}=1.3$ GeV
and $m_0^{K}=1.7$ GeV.
Unit is $10^{-6}$.} 
\label{TABLE2}
\begin{tabular}{|c|ccc|c|c|} \hline
Modes & CLEO & BELLE & BABAR 
& ~~ World Av.~~ &~~~PQCD~~~   \\
\hline  
$\phi K^{\pm}$ & 
 $5.5^{+2.1}_{-1.8}\pm 0.6$ &
 $9.4 \pm 1.1 \pm 0.7$ &  
 $10.0^{+0.9}_{-0.8}\pm 0.5$ &   
 $9.3 \pm 0.8$ &
 $8.1-14.1$  \\
$\phi K^{0}$ & 
 $ 5.4^{+3.7}_{-2.7} \pm 0.7 $ &
 $9.0 \pm 2.2 \pm 0.7$ &  
 $7.6^{+1.3}_{-1.2}\pm 0.5 $ &
 $7.7 \pm 1.1$  &
 $7.6-13.3$    \\ 
\hline
$\phi K^{*\pm}$ & 
 $10.6^{+6.4+1.8}_{-4.9-1.6}$ &  
 $6.7^{2.1+0.7}_{-1.9-1.0} $ & 
 $12.1^{+2.1}_{1.9} \pm 1.1$  &
 $9.4 \pm 1.6$ &
 $12.6-21.2$ \\
$\phi K^{*0}$ & 
 $11.5^{+4.5+1.8}_{-3.7-1.7} $ &
 $10.0^{+1.6+0.7}_{-1.5-0.8} $ &  
 $11.1^{+1.3}_{-1.2}\pm 0.8 $ &
 $10.7 \pm 1.1$    &
 $11.5-19.8$    \\ 
\hline
$K^{*0} \pi^{\pm}$ & 
 $7.6^{+3.5}_{-3.0} \pm 1.6$ &
 $19.4^{+4.2+4.1}_{-3.9-7.1}$ &  
 $15.5 \pm 3.4 \pm 1.8$ &
 $12.3 \pm 2.6$   & 
 $10.2-14.6$  \\
$K^{*\pm}\pi^{\mp}$ & 
 $16^{+6}_{-5} \pm 2 $ &
 $<30$ &  
 $-$ &
 $16 \pm 6$    &
 $8.0-11.6$    \\
$K^{*+} \pi^{0}$ & 
 $<31$ &
 $-$ &  
 $-$ &
 $<31$   & 
 $2.0-5.1 $  \\
$K^{*0}\pi^{0}$ & 
 $<3.6 $ &
 $<7$ &  
 $-$ &
 $<3.6$    &
 $1.8-4.4$  \\   
\hline 
\end{tabular}
\end{table} 

\begin{table}[htb]
\begin{tabular}{|c||c|c||c|c|} \hline 
~~Direct~~$A_{CP}(\%)$~~~~~~& ~~~~~BELLE~~~~~   & ~~~~~BABAR~~~~~   
& ~~~~~~~~~~PQCD~~~~~~~~~~ & ~~~~~~~QCDF~~~~~~ \\ \hline 
$\pi^{+}\pi^{-}$ & $77\pm27\pm8$ & $30 \pm 25 \pm 4$ & 
 $16.0 \sim 30.0$ & $-6\pm12$  \\ \hline
$\pi^{+}\pi^{0}$ & $30\pm30^{+6}_{-4}$ & $-3 \pm 18 \pm 2$ & 
 $0.0$ & 0.0  \\ \hline \hline
$\pi^{+} K^{-}$ & $-6 \pm 9^{+6}_{-2}$ & $-10.2\pm5.0\pm1.6$ & 
$-12.9 \sim -21.9  $ & $5\pm9$   \\ \hline
$\pi^{0}K^{-}$ & $-2\pm19\pm2$& $-9.0 \pm 9.0 \pm1.0$ & 
 $-10.0 \sim -17.3$ & $7\pm9$  \\ \hline
$\pi^{-}\bar{K}^{0}$ & $46\pm15\pm2$ & $-4.7 \pm 13.9$ & 
 $-0.6 \sim -1.5$ & $1\pm1$   \\ \hline 
\end{tabular}
\label{TABLE3}
\caption{CP-asymmetry in $B \to K \pi, \pi\pi $ decays 
with $\phi_3=40^0 \sim 90^0$, $R_b=\sqrt{\rho^2+\eta^2}=0.38$. 
Here we adopted $m_0^{\pi}=1.3$ GeV and $m_0^{K}=1.7$ GeV.}
\end{table} 

\vspace{5mm}
\normalsize
{\bf CP Asymmetry of $B \to \pi\pi, K\pi$:}
Because we have a large imaginary contribution from factorized 
annihilation diagrams in PQCD approach,
we predict large CP asymmetry ($\sim 25 \%$) in $B^0 \to \pi^{+}\pi^{-}$ decays
and about $-15 \%$ CP violation effects in  $B^0 \to K^{+}\pi^{-}$.
The detail prediction is given in Table 3.
The precise measurement of direct CP asymmetry (both magnitude and sign) 
is a crucial way to test factorization models 
which have different sources of strong phases.
Our predictions for CP-asymmetry on $B\to K(\pi)\pi$ have a totally opposite
sign to those of QCD factorization. Recently it was confirmed 
as the first evidence of the direct CP-violation in B-decays that the
DCP asymmetry in $B \to K^{\pm}\pi^{\mp}$ decay is $-0.09 \pm 0.03$ with
$3\sigma$ deviations from zero, 
which is in a good agreement with PQCD result\cite{KLS}. 

\vspace{5mm}
{\bf Radiative B-decays ($B \to (K^*/\rho/\omega) \gamma$):} 
Radiative B-meson decays can provide the most reliable window to understand
the framework of the Standard Model(SM) and to look for New Physics beyond SM
by using the rich sample of B-decays.

In contrast to the inclusive radiative B-decays, exclusive processes such as
$B \to K^{*}\gamma$ are much easier to measure in the experiment with a good
precision\cite{Nakao03}. 
\begin{table}[htb]
\vspace*{0.5cm}
\begin{tabular}{|c||c|c|c|}   \hline \hline
Decay Modes & ~~CLEO~~ & ~~BaBar~~ & ~~Belle~~ 
\\ \hline \hline
{\cal Br}($B \to K^{*0}\gamma $) ($10^{-5}$) &
$4.55 \pm 0.70 \pm 0.34$  & $4.23 \pm 0.40 \pm 0.22$ & 
$4.09 \pm 0.21 \pm 0.19$ \\ \hline
{\cal Br}($B \to K^{*\pm}\gamma $)($10^{-5}$) &
$3.76 \pm 0.86 \pm 0.28 $ & $3.83 \pm 0.62 \pm 0.22$ & 
$4.40 \pm 0.33 \pm 0.24 $ \\ \hline
{\cal Br}($B \to \rho^{0}\gamma $) ($10^{-6}$) &
$< 17$  & $<1.2$ & $< 2.6$ \\ \hline
{\cal Br}($B \to \rho^{+}\gamma $) ($10^{-6}$) &
$< 13 $ & $< 2.1$ & $< 2.7 $ \\ \hline
{\cal Br}($B \to \omega\gamma $) ($10^{-6}$) &
   & $< 1.0$ & $< 4.4 $ \\ \hline \hline
${\cal A}_{CP}(B \to K^{*0}\gamma )$ ($\%$) &
$8 \pm 13 \pm 3$  & $-3.5 \pm 9.4 \pm 2.2$ & 
$-6.1 \pm 5.9 \pm 1.8$  \\ \hline
${\cal A}_{CP}(B \to K^{*+}\gamma )$ ($\%$) &
  &  & $+5.3 \pm 8.3 \pm 1.6$ \\ \hline \hline
\end{tabular}
\caption{Experimental measurements of the averaged branching ratios and 
CP-violating asymmetries of the exclusive $B\to V\gamma$ decays for 
$V=K^{*},\rho$ and $\omega$.
\label{table1}}
\end{table}
\begin{figure}[htb]
\includegraphics[angle=0,width=10.0cm]{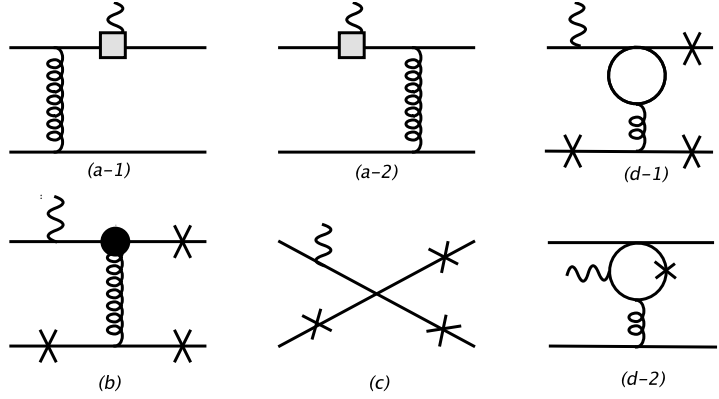} 
\caption{Feynman diagrams of the magnetic penguin(a), chromomagnetic penguin(b),
annihilation(c) and $0_2$-penguin contributions for $B \to V\gamma$ decays}
\label{fig:kstg}
\end{figure}
The main short-distance (SD) contribution to the $B \to K^* \gamma$ decay rate 
involves the matrix element
\begin{equation}
<K^* \gamma| O_{7} |B> ={e m_b \over 8 \pi^2} (-2 i) \epsilon^{\mu}_{\gamma}
<K^* |\bar{s} \sigma_{\mu\nu}q^{\nu}(1-\gamma_5) b | B(p)>,
\end{equation}
which is parameterized in terms of two invariant form fectors as 
\begin{eqnarray}
<K^*(P_3,\epsilon_3)|\bar{s} \sigma_{\mu\nu}q^{\nu}(1-\gamma_5) b | B(P)>
&=& [\epsilon_{3,\mu}(q\cdot P)-P_{\mu}(q\cdot\epsilon_3)] \cdot 2 T_2(q^2) 
\nonumber \\
&& \hspace{15mm} +i\epsilon_{\mu\nu\alpha\beta} \epsilon_3^{\nu} P^{\alpha}
q^{\beta} \cdot 2 T_1(q^2).
\end{eqnarray}
Here $P$ and $P_3=P-q$ are the B-meson and $K^*$ meson momentum, respectively
and $\epsilon_3$ is the polarization vector of the $K^*$ meson.
For the real photon emission process the two form factors coincide,
$T_1(0)=T_2(0)=T(0)$. 
This form factor can be calculable in the $k_T$ factorization
method including the sudakov suppression factor and 
the threshold resummation effects. As discussed in ref\cite{KM03}, 
we obtain $T(0)=0.28 \pm 0.02 $ for $B \to K^* \gamma$ 
which is far away from 
the QCD result $0.38\pm 0.06$ by using the light-cone QCD sum rule \cite{BB-98}, 
however in accordance with the preliminary result of Lattice QCD, 
$0.25 \pm 0.06$\cite{Beci-03}.

Even though theoretical predictions for the exclusive decays always has
large model dependent hadronic uncertainties, such uncertainties can be 
cancelled in the searching of the CP-asymmetry and the isospin breaking effect.

Including all possible contributions from $0_{7\gamma},0_{8g},0_2$-penguin
and annihilation in Figure 2, we obtain 
the Branching ratios:
\begin{itemize}
\item $Br(B^0 \to K^{0*}\gamma) = 
( 3.5^{+1.1}_{-0.8} ) \times 10^{-5} $ \hspace{7mm}
$Br(B^+ \to K^{+*}\gamma) = (3.4^{+1.2}_{-0.9}) \times 10^{-5}, $
\item $Br(B^0 \to \rho^{0}\gamma) = (0.95 \pm 0.07) \times 10^{-6} $ \hspace{5mm}
$Br(B^+ \to \rho^{+}\gamma) = (1.63 \pm 0.20) \times 10^{-6} $, 
\end{itemize}
and the CP-Asymmetry :
\begin{itemize}
\item $Acp(B^0 \to K^{0*}\gamma) = ( 0.39^{+0.06}_{-0.07}) \% $ \hspace{10mm}
$Acp(B^+ \to K^{+*}\gamma) = (0.62 \pm 0.13)  \%$
\end{itemize}
The small difference in the branching fraction between $K^{0*}\gamma$
and $K^{+*}\gamma$ can be detected as the isopsin symmetry breaking  
which tells us the sign of the combination of the Wilson coefficients, $C_6/c_7$.
We obtain
\begin{equation}
\Delta_{0-}={\eta_{\tau} Br(B \to \bar{K}^{0*}\gamma) - Br(B \to K^{*-}\gamma)
\over \eta_{\tau} Br(B \to \bar{K}^{0*}\gamma) + Br(B \to K^{*-}\gamma) }
= (5.7^{+1.1}_{-1.3}  \pm 0.8  ) \%
\end{equation}
where $\eta_{\tau}=\tau_{B^+}/\tau_{B^0}$. The first error term comes from
the uncertainty of shape parameter of the B-meson wave function 
($0.38 < \omega_B < 0.42$)
and the second term is origined from the uncertainty of $\eta_{\tau}$. 
By using the world averaged value of measurement and 
$\tau_{B^+}/\tau_{B^0}=1.083 \pm 0.017$, we find numerically that
$\Delta_{0-}(K^*\gamma)^{exp}=(3.9 \pm 4.8) \%$. In PQCD we can not expect
large isospin symmetry breaking in $B \to K^* \gamma$ system.

\section{Extraction of $\phi_2(=\alpha)$ from $B \to \pi^{+}\pi^{-}$}
Even though isospin analysis of $B \to \pi\pi$ can provide a clean way
to determine $\phi_2$, it might be difficult in practice because of
the small branching ratio of $B^0 \to \pi^0\pi^0$.
In reality in order to determine $\phi_2$, we can use the time-dependent rate
of $B^0(t) \to \pi^{+}\pi^{-}$.
Since penguin contributions are sizable about 20-30 \% of the total amplitude,
we expect that direct CP violation can be large if strong phases are different
in the tree and penguin diagrams.

In our analysis we use the c-convention.
The ratio between penguin and tree amplitudes is $R_c=|P_c/T_c|$ 
and the strong phase difference
between penguin and tree amplitudes $\delta=\delta_P-\delta_T$.
The time-dependent asymmetry measurement provides two equations for
$C_{\pi\pi}$ and $S_{\pi\pi}$ in terms of three unknown variables 
$R_c,\delta$ and $\phi_2$\cite{GR}.
\begin{figure}[t!]
\includegraphics[angle=-90,width=0.6\textwidth]{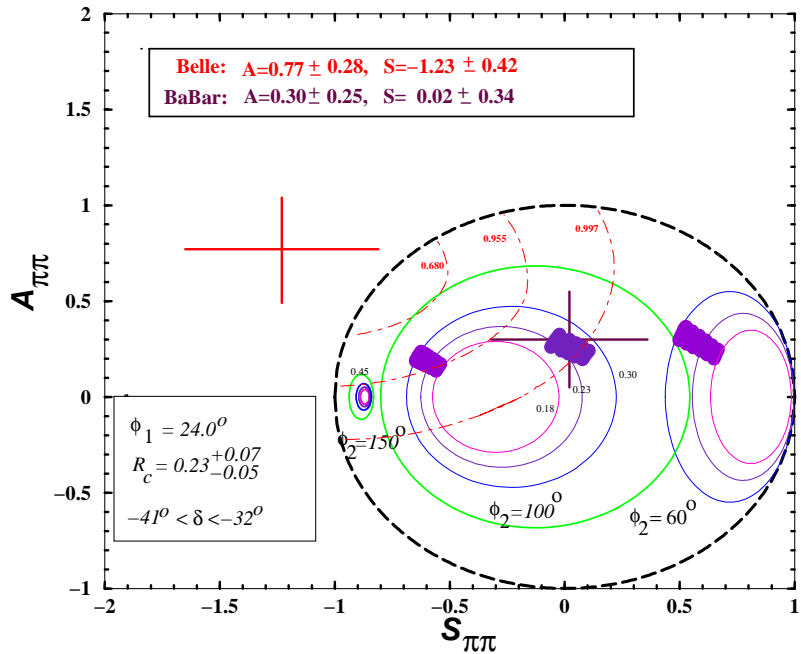} 
\caption{Plot of $A_{\pi\pi}$ versus $S_{\pi\pi}$  for various values
of $\phi_2$ with $\phi_1=24.3^o$, $0.18 < R_c < 0.30$ and $-41^o <
\delta < -32^o$ in the pQCD method.}
\label{fig:cpipi}
\end{figure}
Since pQCD provides us $R_c=0.23^{+0.07}_{-0.05}$ and $-41^o
<\delta<-32^o$, the allowed range of $\phi_2$ at present stage is
determined as $55^o <\phi_2< 100^o$ as shown in Figure \ref{fig:cpipi}. 

According to the power counting rule in the pQCD approach,
the factorizable annihilation contribution with large imaginary part
becomes subdominant and give a negative strong phase from 
$-i\pi\delta(k_{\perp}^2-x\,M_B^2)$.
Therefore we have a relatively large
strong phase in contrast to QCD-factorization ($\delta\sim 0^o$) 
and predict large direct CP violation effect 
in $B^0\to \pi^{+}\pi^{-}$ 
with $A_{cp}(B^0 \to \pi^{+}\pi^{-}) = (23\pm7) \%$, 
which will be tested by more precise experimental measurement within two years. 

In the numerical analysis, since the data by Belle
collaboration\cite{belle} 
is located ourside allowed physical regions, we considered the
recent BaBar measurement\cite{babar} with $90\%$ C.L. interval
taking into account the systematic errors:
\begin{itemize}
\item[$\bullet$]
$S_{\pi\pi}=  0.02\pm0.34\pm0.05$ 
\hspace{15mm} [-0.54,\hspace{5mm} +0.58]
\item[$\bullet$]
$A_{\pi\pi}= 0.30\pm0.25\pm0.04$ 
\hspace{14mm} [-0.72,\hspace{5mm} +0.12].
\end{itemize}
The central point of BaBar data corresponds to $\phi_2 = 78^o$ 
in the pQCD method. 
Even if the data by Belle
collaboration\cite{belle} 
is located ourside allowed physical regions, we can have allowed
ranges with 2 $\sigma$ bounds, but large negative $\delta$ and 
$R_c > 0.4$ is prefered\cite{KS-ckm03}.
\begin{figure}[ht]
\caption{Allowed ranges of $(\Delta_{LR})_{23}$ and plot
of the correlation between $S_{\phi K_s}$ and $C_{\phi K_s}$} 
\includegraphics[angle=-90,width=0.4\textwidth]{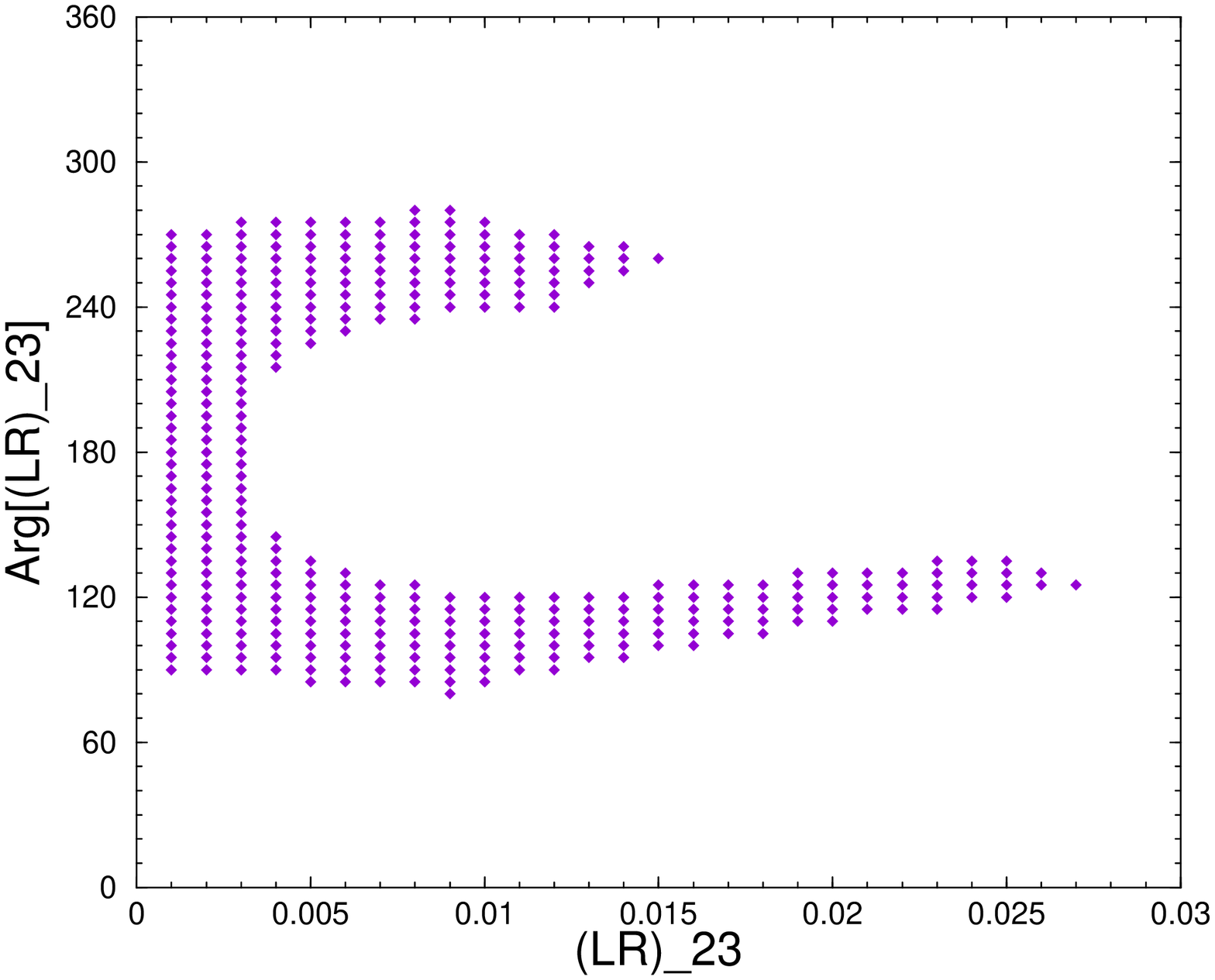} 
\hspace{1.0cm}
\includegraphics[angle=-90,width=0.4\textwidth]{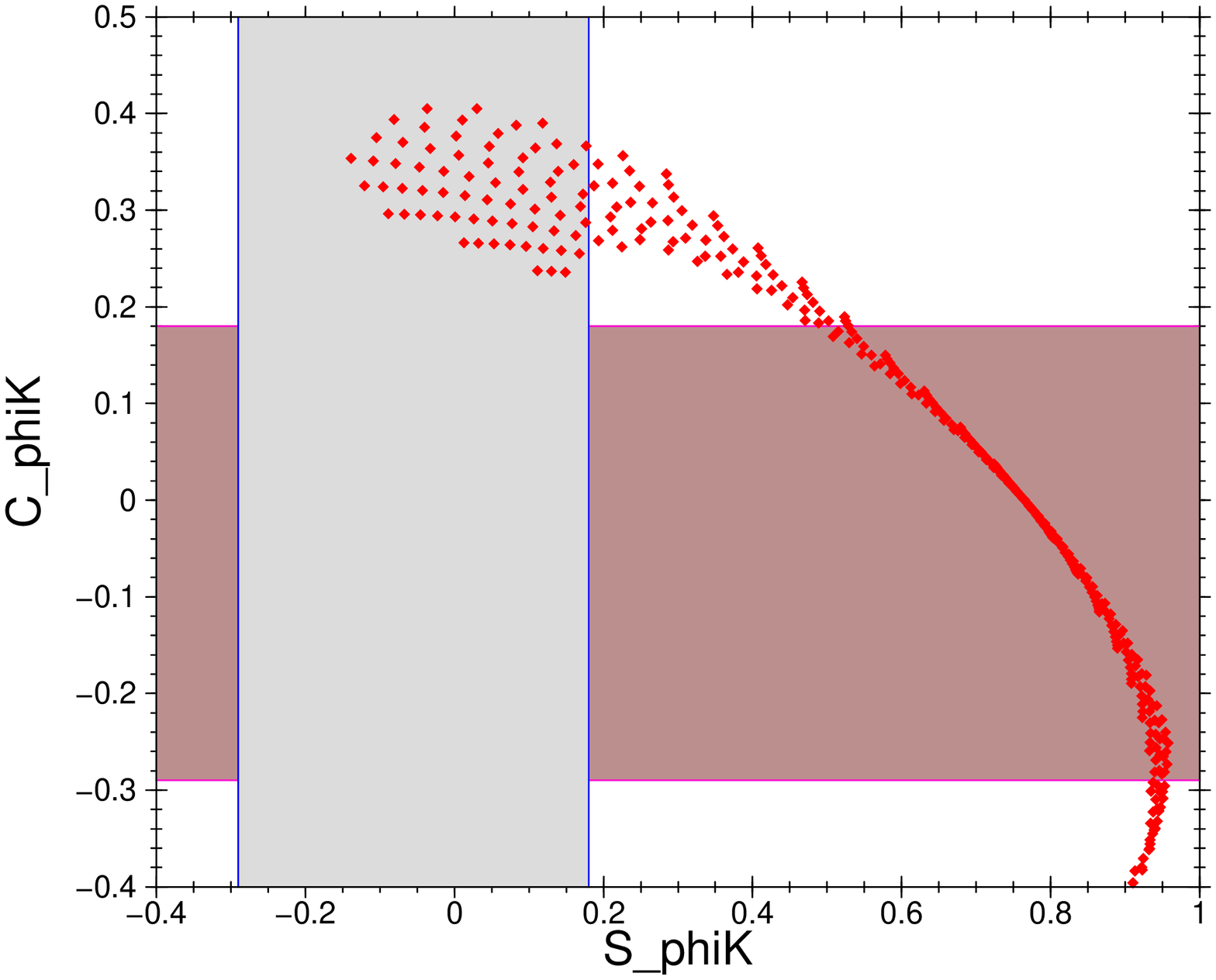} 
\end{figure}
\section{New Physics Search in $B \to \phi K_s$ decays}
In SM the time-dependent CP asymmetry in $B \to \phi K_s$ 
is expected the same as one in $B \to J/\psi K_s$, 
$sin(2\phi_1)(J/\psi K_s)=0.734 \pm 0.054$. Recently Belle measured
$S_{\phi K_s} = -0.96 \pm 0.50^{+0.09}_{-0.11}$\cite{Belle-phiks} 
and BaBar obtained $0.45 \pm 0.43 \pm 0.07$\cite{BaBar-phiks}. 
The world averaged value $-0.15 \pm 0.33$
with 2.7 $\sigma$ deviation from SM prediction shows large possibilities
of new physics contributions in the decay amplitue of $B \to \phi K_s$
though the quantum loop effect. We consider the new physics contribution
of gluino mediation SUSY in the MSSM. Among four possible contributions
($LL,RR,LR$ and $RL$-insertions), the LR and RL-contributions can be dominant,
while the LL and RR-contribution are suppressed strongly from 
$B_s - \bar{B}_s$ mixing. Here we show the results of LR-insertion 
within PQCD approach including vertex corrections. 
In this case, the analysis can be consistent because the amplitudes of 
SM and new physics part keeps upto $0(\alpha_s^2)$ terms in the short distance part.

In our numerical analysis we used the following constraints:
\begin{itemize}
\item $2.0 \times 10^{-4} < Br(b \to s \gamma) < 4.5 \times 10^{-4}$, 
\hspace{10mm} $-27 \% < A_{cp}(b \to s \gamma) < 10 \%$,
\item $Br(B^0 \to X_s l^{+} l^{-}) = (6.1\pm 1.4 \pm 1.3) \times 10^{-6}$,
\hspace{10mm} $\Delta M_s > 14.4 ps^{-1}$.
\end{itemize}
In LR-insertion case, $C_{8g}$ contributions can be important both
for the branching ratio and the CP-asymmetry and the most strong constraint
comes from $Br(B \to X_s \gamma)$. As shown in figure 4, 
$S_{\phi Ks}$ can be reach to $-20 \%$ and $C_{\phi Ks} < 40 \%$.  
The detail analysis will appear elsewhere\cite{YYK-susyLR}.
\section{Summary and Outlook}
In this talk we have discussed ingredients of PQCD approach and some important
theoretical issues with numerical results by comparing exparimental data.
The PQCD factorization approach provides a useful theoretical framework
for a systematic analysis on non-leptonic two-body B-meson decays including
radiative decays. Our results are in a good agreement with experimental data. 
Specially pQCD predicted large direct CP asymmetries
in $B^0 \to \pi^{+}\pi^{-}, K^{+}\pi^{-}$ decays, 
which will be a crucial touch stone to distinguish our approach 
from others in future precise measurement. Recently the measurement
of the direct CP asymmetry in $B\to K^{\pm}\pi^{\mp}$, 
$A_{cp}(K^{+}\pi^{-})=-9\pm 3 \%$ is in accordance with our prediction.
For other decay modes, for instance $B \to D^{(*)} \pi$\cite{KLKLS}, 
H.N.Li has summarized it in this conference.    

We discussed the method to determine weak phases 
$\phi_2$ within the pQCD approach 
through Time-dependent asymmetries in $B^0\to
\pi^{+}\pi^{-}$. 
We get interesting bounds on $55^o < \phi_2 < 100^o$ with
90\% C.L. of the recent BaBar measurement.

For the time-dependent CP-asymmetry of $B \to \phi K_s$, we also explore
the possibility of the new physics contributions from gluino mediation SUSY in MSSM.
\begin{center}{\bf\Large Acknowledgments}
\end{center}
It is a great pleasure to thank E.J.Chun for the invitation
at this exciting conference on flavor physics at KIAS in Korea.
We wish to acknowlege the fruitful collaboration with H.-N. Li 
and joyful discussions with other members of PQCD working group.
This work was supported in part by the Japan Society for
the Promotion of Science and in part
by Grant-in Aid for the 21st Century COE-program at Nagoya University.



\begin{thebibliography}{99}
\bibitem{BSW} M. Bauer, B. Stech, M. Wirbel,
Z. Phys. C {\bf 29}, 637 (1985); {\sl ibid.} {\bf 34}, 103 (1987).
\bibitem{Cheng94} H.Y. Cheng, Phys. Lett. B {\bf 335}, 428 (1994).
\bibitem{Cheng96} H.Y. Cheng, Z. Phys. C {\bf 69}, 647 (1996).
\bibitem{Soares} J. Soares, Phys. Rev D {\bf 51}, 3518 (1995);
A.N. Kamal and A.B. Santra, Z. Phys. C {\bf 72}, 91 (1996);
A.N. Kamal, A.B. Santra, and R.C. Verma, Phys. Rev. D {\bf 53},
2506 (1996).
\bibitem{BBNS}  M. Beneke, G. Buchalla, M. Neubert, and C.T. Sachrajda,
Phys. Rev. Lett. {\bf 83}, 1914 (1999);
Nucl. Phys. {\bf B591}, 313 (2000).
\bibitem{LB} G.P. Lapage and S.J. Brodsky, Phys. Lett. B {\bf 87}, 359
(1979); Phys. Rev. D {\bf 22}, 2157 (1980).
\bibitem{BS} J. Botts and G. Sterman, Nucl. Phys. {\bf B225}, 62 (1989).
\bibitem{LS} H-n. Li and G. Sterman, Nucl. Phys. {\bf B381}, 129
(1992).
\bibitem{LY1} H-n. Li and H.L. Yu, Phys. Rev. Lett. {\bf 74}, 4388 (1995);
Phys. Lett. B {\bf 353}, 301 (1995); Phys. Rev. D {\bf 53}, 2480 (1996).
\bibitem{CL} C.H. Chang and H-N. Li, Phys. Rev. D {\bf 55}, 5577 (1997).
\bibitem{YL} T.W. Yeh and H-N. Li, Phys. Rev. D {\bf 56}, 1615 (1997).
\bibitem{CLY} H.Y. Cheng, H-N. Li, and K.C. Yang,
Phys. Rev. D {\bf 60}, 094005 (1999).
\bibitem{NL} H-N. Li, Phys. Rev. D {\bf 64}, 014019 (2001); M. Nagashima
and H-N. Li, hep-ph/0202127; Phys. Rev. D {\bf 67}, 034001 (2003).
\bibitem{KLS} Y.-Y. Keum, H-N. Li, and A.I. Sanda,
Phys. Lett. B {\bf 504}, 6 (2001); Phys. Rev. D {\bf 63}, 054008 (2001);
Y.Y. Keum and H-n. Li, Phys. Rev. {\bf D63}, 074006 (2001).
\bibitem{LUY} C. D. L\"{u}, K. Ukai, and M. Z. Yang,
Phys. Rev. D {\bf 63}, 074009 (2001).
\bibitem{KS} Y.-Y. Keum and A. I. Sanda, Phys. Rev. D {\bf 67}, 054009
(2003).
\bibitem{Keum02} Y.-Y. Keum, hep-ph/0210127.
\bibitem{Keum01} Y.-Y. Keum, hep-ph/0209208;hep-ph/0209002;
M. Battaglia et al.,Future Directions in the CKM matrix and the Unitarity Triangle,
hep-ph/0304132.
\bibitem{Bro} G.P.~Lepage and S.J.~Brodsky, Phys. Rev. {\bf D22}, 2157
(1980).
\bibitem{BSH}
A.~Szczepaniak, E.M.~Henley and S.~Brodsky, \emph{Phys. Lett. {\bf B}}
  \textbf{243}, 287 (1990).

\bibitem{CKL:a}
C.-H.~Chen, Y.-Y.~Keum  and H.-N.~Li, \emph{Phys. Rev. {\bf D}}
  \textbf{64}, 112002 (2001).

\bibitem{Kurimoto}
T.~Kurimoto, H.-N.~Li and A.I.~Sanda, \emph{Phys. Rev. {\bf D}}
  \textbf{65}, 014007 (2002).
\bibitem{BSS}
M.~Bander, D.~Silverman and A.~Soni, \emph{Phys. Rev. Lett.} 
\textbf{43}, 242 (1979).

\bibitem{CK}
H.-Y.~Cheng and K.-C.~Yang, \emph{Phys. Rev. {\bf D}} 
\textbf{64}, 074004  (2001).
H.-Y.~Cheng Y.-Y.~Keum and K.-C.~Yang, \emph{Phys. Rev. {\bf D}} 
\textbf{65}, 094023  (2002).
 
\bibitem{BBNS:99}
M.~Beneke, G.~Buchalla, M.~Neubert and C.T.~Sachrajda,  \emph{Phys. Rev.
  Lett.} \textbf{83}, 1914 (1999).

\bibitem{Mishima}
S.~Mishima, \emph{Phys. Lett. {\bf B}} \textbf{521}, 252 (2001).
\bibitem{CKL:b}
C.-H.~Chen, Y.-Y.~Keum  and H.-N.~Li, \emph{Phys. Rev. {\bf D}} 
\textbf{66}, 054013  (2002).

\bibitem{Nakao03}M. Nakao,Proceedings of Lepton-Photon `03 conference [hep-ex/0312041].
\bibitem{BB-98}P. Ball and V.M. Braun, Phys. Rev. D{\bf 58}, 094016 (1998).
\bibitem{Beci-03}D. Becirevic, talk given at the Flavour Physics and CP violation,
Paris, France, May 2003;hep-ph/0211340.

\bibitem{KM03} Y.-Y.~Keum, M.~Matsumori, and A.I.~Sanda, 
CP-Asymmetry, Branching ratios in $B \to V\gamma$ within
$k_T$ factorization, hep-ph/0406055.

\bibitem{GR}
R.~Fleischer and J.~Matias, Phys.Rev.{\bf D66} (2002) 054009  ; 
M.~Gronau and J.L.~Rosner, Phys.Rev.{\bf D65} (2002) 013004,
Erratum-ibid.{\bf D65} (2002) 079901;
Phys. Rev. {\bf D65} (2002) 093012; hepph/0205323.
\bibitem{belle}
Belle Collaboration (K. Abe et al.), Belle-preprint 2002-8 [hep-ex/0204002].
\bibitem{babar}
BaBar Collaboration (B. Aubert et al.), BaBar-Pub-02/09 [hep-ex/0207055].
\bibitem{KS-ckm03}Y.-Y. Keum and A.I.Sanda, eConf C0304052: WG420,2003
[hep-ph/0306004].
\bibitem{Belle-phiks} 
Belle Collaboration (K. Abe et al.), Belle-preprint 2003-14 [hep-ex/0308035].
\bibitem{BaBar-phiks}
T. Browder,CKM phases($\beta/\phi_1$), 
Talk presented at Lepton-Photon 2003,hep-ex/0312024.
\bibitem{YYK-susyLR} Y.-Y. ~Keum, Vertex corrections and 
New Physics Search in $B \to \phi K_s$
within the PQCD approach, to appear.
\bibitem{KLKLS}Y.-Y. Keum, T. Kurimoto, H.-n. Li, C.-D. Lu, and A.I. Sanda,
Phys. Rev. {\bf D69};094018,2004;[hep-ph/0305335].
\end{thebibliography}
\end{document}